\documentstyle[aps,prl,balanced,times,psfig]{revtex}
\begin{document}
\title{COOPERATIVE PARRONDO'S GAMES}
\author{Ra\'ul Toral}
\address{Instituto Mediterr\'aneo de Estudios Avanzados, IMEDEA (CSIC-UIB)\\ Campus UIB, 07071-Palma de Mallorca, Spain,\\
\footnotesize\it email: raul@imedea.uib.es,  http://www.imedea.uib.es/PhysDept} 

\maketitle
\begin{abstract}We introduce a new family of Parrondo's games of alternating losing strategies in order to get a winning result. In our version of the games we consider an ensemble of players and use ``social" rules in which the probabilities of the games are defined in terms of the actual state of the neighbors of a given player.\end{abstract}
\vspace{1.0cm}
%\begin{twocolumns}

Parrondo's paradox\cite{ha99,ha99b,hatp00,hatpp00} shows that the combination of two losing strategies can lead to a winning result. The paradox can be phrased in terms of very simple gambling games in which some unit (say, 1 euro) is won or lost with a given probability. We can imagine that the games consist on tossing different biased coins and that a ``capital" $C(t)$ is built. Every time a game is played (a coin is tossed) time increases by one unit, $t\to t+1$, and the capital increases or decreases by $1$, $C(t)\to C(t)\pm 1$. Games are classified as winning, losing or fair if the average capital $\langle C(t)\rangle$ increases, decreases or remains constant with time, respectively.

The original version of the paradox is based on the two following basic games that can be played at any time $t$:\\
\indent $\bullet$ Game A: there is a probability $p$ of winning.\\
\indent $\bullet$ Game B: If the capital $C(t)$ is a multiple of $3$, the probability of winning is $p_1$, otherwise, the probability of winning is $p_2$.\\
Game A is fair if $p=1/2$ and it is easy to prove that game B is fair if the condition $(1-p_1)(1-p_2)^2=p_1p_2^2$ holds. Choosing, for example, $p=0.5-\epsilon$, $p_1=0.1-\epsilon$, $p_2=0.75-\epsilon$, with $\epsilon$ a small positive number it turns out that both game A and game B, when played by themselves, are losing games. The surprise arises when game A and B are played alternatively, either in succession such as AABBAABBAABB$\dots$  or randomly by choosing (with probability $1/2$) the next game to be played. We will denote, for short, game ``A+B"  the case in which games A or B are chosen randomly at each time step and game ``$[2,2]$" corresponds to the sequence AABBAABBAABB$\dots$ In these cases of alternation in the game played, the paradoxical result is that the outcome turns out to be a winning game, although both games A and B are losing games. The paradox can be related to a discrete version of the flashing ratchet\cite{ab94} in which a Brownian particle is subjected, alternatively, to the effect of two potentials. The potentials are such that each one, acting alone, induces a net movement for the particle in the direction of decreasing potential (say, to the left). However, in the case of alternating potentials, the net movement is in the opposite direction (to the right). The possible relevance of the paradox to other situations of interest in fields such as Biology, Economy and Physics has been recently pointed out\cite{pha00,d01}.

A new version of the paradox has been introduced recently\cite{pha00}. In this new version, game A is the same as defined before, but the probabilities of game B do not depend directly on the actual value of the capital, but rather on the past history of winning and losing games. Let us classify the game played at any time $t$ as a ``loser" (``winner") if it results in loss (win) of capital. The probabilities of game B depend then on the outcome of the last two games. More precisely, game B is modified as follows: the probability of winning at time $t$ is:\\
\indent $\bullet$ $p_1$, if game at $t-2$ was loser and game at $t-1$ was loser,\\
\indent $\bullet$ $p_2$, if game at $t-2$ was loser and game at $t-1$ was winner,\\
\indent $\bullet$ $p_3$, if game at $t-2$ was winner and game at $t-1$ was loser,\\
\indent $\bullet$ $p_4$, if game at $t-2$ was winner and game at $t-1$ was winner.\\
The condition for the new game B to be a fair one is $p_1p_2=(1-p_3)(1-p_4)$. For instance, the choice $p=1/2-\epsilon, p_1=0.9-\epsilon$, $p_2=p_3=0.25-\epsilon$, $p_4=0.7-\epsilon$ (for positive $\epsilon$) results in A and B being both losing games. However, it has been shown that (for sufficiently small $\epsilon$) the paradox appears again: the alternation games A+B or $[2,2]$ are both winning games.

Inspired by this result, we introduce a new version of the games that we call
``cooperative Parrondo's games". In this version, we consider an ensemble of
$N$ players, each one owning a capital $C_i(t)$, $i=1,\dots,N$. This capital
evolves by combination of two basic games, A and B. First a player, $i$, is
randomly chosen between the $N$ players. Player $i$ can then play either game
A or game B according to some rules. When a player wins in any particular
game, it is labeled as ``winner", otherwise it is labeled as ``loser". The
state of ``winner" or ``loser" for a particular player remains unchanged in
time until that player has the chance to play again. Game A is the same as
before: there is a probability $p$ (alternatively $1-p$) for player $i$ to win
(lose) one unit and increase (decrease) his capital $C_i(t)$ by $1$. The new
ingredient is that the probabilities of  game B are defined according to the
state of the two neighbor players, $i+1$ and $i-1$ (we assume periodic boundary
conditions). The simplest version appears when the probability of winning at
time $t$ depends on the present state of the neighbors and it is given by:\\
\indent $\bullet$ $p_1$, if player at site $i-1$ is a loser and player at site $i+1$ is a loser,\\
\indent $\bullet$ $p_2$, if player at site $i-1$ is a loser and player at site $i+1$ is a winner,\\
\indent $\bullet$ $p_3$, if player at site $i-1$ is a winner and player at site $i+1$ is a loser,\\
\indent $\bullet$ $p_4$, if player at site $i-1$ is a winner and player at site $i+1$ is a winner.\\
We classify the games according to the behavior of the total capital $C(t)=\sum_i C_i(t)$. Therefore, a winning game is one for which the average value of the total capital $\langle C(t)\rangle$ increases with time, and similarly for losing and fair games. 

We will show that Parrondo's paradox applies also to this kind of games: there
are situations, i.e. values of the probabilities $p,p_1,p_2,p_3,p_4$, in which
the alternation of game A and game B can perform better that games A and
B separately. In particular, it is possible for games A and B to be
losing or fair  games, and that an alternation of games A and B leads to a
winning result. Before presenting numerical evidence of this paradoxical
result, we perform a mean-field type analysis that can capture some aspects of
this new game. 

Let $P^A_i(t)$, $P^B_i(t)$ and $P^{A+B}_i(t)$ be the probability
that the $i$ player is labeled as winner at time $t$ when playing only game
A, only game B or game A+B, respectively. For game A it is
$P^A_i(t)=p$ for all times $t$. For games B and A+B the evolution of the probabilities is
governed by master equations which can not be solved explicitly. In our
mean-field approach we assume that the probabilities of being winner or loser
at time $t$ for the different players are equal for all of them, i.e.
$P^B_i(t)\equiv P^B(t)$ and $P^{AB}_i(t)\equiv P^{AB}(t)$. The mean-field
equation for the evolution of the common probability in game B is the map:
\begin{equation}
P^B(t+1)  =  (1-P^B(t))^2p_1+P^B(t)(1-P^B(t))(p_2+p_3)+P^B(t)^2p_4
\equiv  F(P^B(t)),
\end{equation}
where the different terms of this equation reflect each of the $4$ possibilities that have been spelt out in the definition of game B.

In the random sequence defining game A+B, there is a probability $1/2$ of choosing game A and a probability $1/2$ of choosing game B at each time step. Therefore the corresponding mean-field equation for game A+B is the map $P^{A+B}(t+1)=\frac{1}{2}\left[p+ F(P^{A+B}(t))\right]$.
The stationary probability of winning in game B is obtained by setting $P^B(t+1)=P^B(t)\equiv P^B$ and satisfies the equation 
$P^B=F(P^B)$.
Similarly, the mean field stationary solution $P^{A+B}$ for game A+B is obtained by solving $P^{A+B}=\frac{1}{2}[p+F(P^{A+B})]$. As for any quadratic map, there is at most one stable stationary solution of the maps of game $B$ or $A+B$. It can be shown that for game $B$ there is a stable solution $P^{B}\in[0,1]$ for $p_4\ge 1/4$, whereas for $p_4<1/4$, there are no stable solutions for some values of the set of probabilities $\{p_1,p_2,p_3\}$ defining the game. On the other hand, there is always a stable solution for the map $A+B$ such that $P^{A+B}\in[0,1]$.

We will find a ``weak" paradoxical result of the Parrondo type whenever the random alternation of game A+B performs better than the average of games A and B. This is to say, whenever the condition $
P^{A+B}>\frac{1}{2}(p+P^B)$
is satisfied. This depends on the set of numbers $p,~p_1,~p_2,~p_3,~p_4$. For instance, choosing $p=0.5,\quad p_2=p_3=0.16,\quad p_4=0.7$ the inequality is satisfied both for $p_1<0.5395$ and for $p_1> 0.98$\footnote{These numbers by no means constitute an exceptional example. Many other sets of probabilities lead to this paradoxical result.}.

\begin{figure}[!t] 
\centerline{\psfig{file=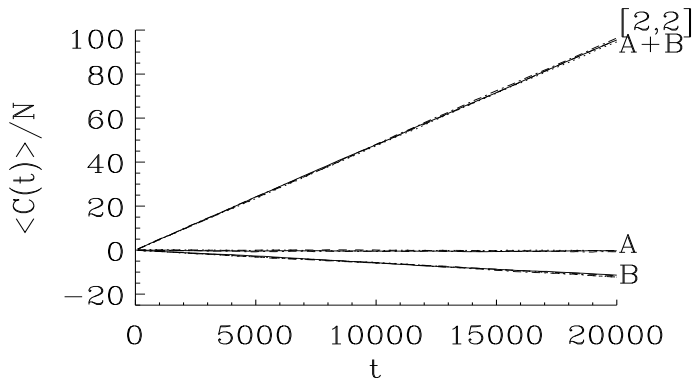,width=10.0cm,height=5.0cm}} 
%\vspace*{6pt}
\caption{Average capital per player, $\langle C(t)\rangle/N$, versus time, $t$, measured in units of games per player. The probabilities defining the games are: $p=0.5,p_1=1,p_2=p_3=0.16,p_4=0.7$. These results show that game A is fair, game B is a losing game, but that when games A are B are played in random succession (game A+B) or in the $[2,2]$ alternation AABBAABB$\dots$, the result is a winning game. We show in this graph results for $N=50,~100,~200$ players. Averages are performed with respect to $M=1000$ realizations of the games using different random numbers.
\label{fig1}} 
\end{figure}

We now describe the results obtained after computer simulations of the model.
The numerical results, in accordance with the mean field analysis, do show the
existence of sets of probabilities $p,~p_1,~p_2,~p_3,~p_4$ for which the game
A+B performs better than the average of games A and B played separately (``weak
paradoxical result"). This happens, for instance, for
$p=0.5,p_1=0.5,p_2=p_3=0.16,p_4=0.7$ in which both games B and game A+B are
losing games. More interestingly, there are situations in which the paradoxical
result is enhanced by the fact that game B is a losing game whereas game A+B
turns out to be a winning game. This ``strong" paradoxical result, which is
observed in the numerical simulation, is not correctly predicted by the
mean-field analysis\footnote{This statement is based upon a numerical comparison
of the mean field solutions. This analysis does not exclude completely that a
strong paradoxical result can be obtained within the mean field theory in a tiny
region of parameter space $\{p,p_1,p_2,p_3,p_4\}$, but we believe this is a very unlikely possibility.}.  As a way of example, we
plot in  figure \ref{fig1} the average capital when playing the games using the
values $p=0.5,p_1=1,p_2=p_3=0.16,p_4=0.7$. As shown in the figure, and for
these values of the probabilities, it turns out that game A is fair, game B is
losing, but both A+B and $[2,2]$ are winning games. Mean field in this case, yields $P^{B}=0.50386$, $P^{A+B}=0.50218$, and it is able to predict only a weak paradoxical result: $P^{A+B}> (p+P^B)/2$, but can not predict the strong paradoxical result $P^{A+B}>P^B$, as observed in the simulation. By choosing
$p=0.5-\epsilon$, for small enough $\epsilon$, it is possible to turn game $A$
into a losing game while still keeping game $A+B$ as a winning game.  The
existence of this paradoxical result when playing games whose probabilities
depend on the present state of the neighbors is the main result of this paper.

\begin{figure}[!t]
\centerline{\psfig{file=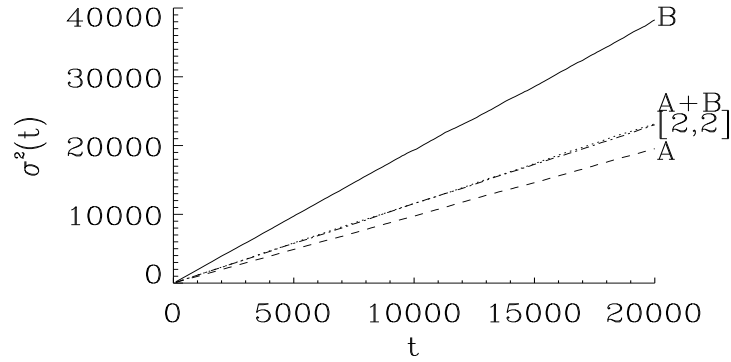,width=10.0cm,height=5.0cm}}
%\vspace*{6.0pt}
\caption{Time evolution of the variance $\sigma^2$ of the single player capital distribution using games with the same probabilities than in figure \ref{fig1} and $N=50$ players. For game A the exact result is $\sigma^2=t$. A linear fit to the law $\sigma^2=\alpha t$ gives $\alpha= 1.91 $ for game B and $\alpha= 1.15$ for both games A+B and $[2,2]$. This indicates that the spread of capital is less in the combined games than in the single game B.
\label{fig2}}
\end{figure}

We next characterize the capital distribution between the different players. We
expect, due to the law or large numbers, that the probability distribution
$P(C_i,t)$ of the capital $C_i$ of player number $i$ at time $t$ follows in all
cases, and in the limit of large $N$, a Gaussian distribution. Due to the
symmetry between players, the average capital of a
single player is equal for all of them, $\langle C_i(t) \rangle= N^{-1}\langle
C(t) \rangle$. We have computed the variance of the single-player capital
distribution $\sigma^2(t)=\langle N^{-1}\sum_i C_i(t)^2-[\langle C(t)\rangle/N]^2 \rangle$.
For game A we have the exact result: $N^{-1}\langle C(t)\rangle=(2p-1)t$ and
$\sigma^2(t)=4p(1-p)t$. We can see in figure \ref{fig2} that the  variance of
game B is, for the particular values of the probabilities taken here, larger
than that of game A. However, the random alternation of game A+B or the
sequence $[2,2]$ both yield a similar variance  which is smaller than
that of game B. We conclude that the distribution of capital between the
different players for the combined games A+B or $[2,2]$ is more uniform than
that of game B. This interesting result excludes that a winning overall game
is produced as a consequence of a large fluctuation implying only a small
number of players.

In conclusion, we have introduced a new family of games based upon the Parrondo's strategy of alternating non-winning games to generate a winning result. The novelty of our treatment lies on that a winning game results from the interaction between the different players. In that sense, we can consider that our results can serve as a model for the application of the games to situations in which ``social" rules for the definition of the success probabilities need to be considered. Extensions of the games in which the players are connected by more realistic small-world type networks can also be considered.

\vspace{5pt}
\noindent{\bf Acknowledgements:} This work was supported by MCYT (Spain), grants numbers PB97-0141-C01-01 and BFM2000-1108.

%\end{twocolumns}
\end{document}